\documentclass[12pt]{article}
\def\a{\alpha}

\usepackage{ae}
\usepackage[T1]{fontenc}
\usepackage[ansinew]{inputenc}
\usepackage{mathrsfs}
\usepackage{amsmath}
\usepackage{amssymb}
\usepackage{graphicx}
\usepackage{color}
\definecolor{darkblue}{cmyk}{0.9,0.9,0,0}
\definecolor{carageen}{RGB}{0,0.55,0}
\usepackage[colorlinks=true,linkcolor=darkblue,citecolor=darkblue,urlcolor=darkblue]{hyperref}
\usepackage{epsfig}
\usepackage{wasysym}
\usepackage{MnSymbol}

\usepackage{amsmath,amsfonts,amssymb,amsbsy}
\usepackage{graphicx,color}
\usepackage{graphics}
\usepackage{epsfig}
\usepackage{verbatim}
\usepackage{stmaryrd}
\usepackage{epsf}
\usepackage{amsmath, amssymb,graphicx}
\usepackage{epsfig}
\usepackage{verbatim}
\usepackage{amsfonts}
\usepackage{ulem}
\usepackage{cite}

\def\bc{\begin{center}}
\def\ec{\end{center}}

\newcommand{\be}{\begin{equation}}
\newcommand{\ee}{\end{equation}}
\newcommand{\ba}{\begin{eqnarray}}
\newcommand{\ea}{\end{eqnarray}}
\newcommand{\nn}{{\nonumber}}
\newcommand{\beaa}{\begin{eqnarray}}
\newcommand{\eeaa}{\end{eqnarray}}

\definecolor{carageen}{RGB}{0.1,0.7,0.1}

\DeclareFontFamily{OT1}{pzc}{}
\DeclareFontShape{OT1}{pzc}{m}{it}{<-> s * [1.10] pzcmi7t}{}
\DeclareMathAlphabet{\mathpzc}{OT1}{pzc}{m}{it}

\def\({\left(}
\def\){\right)}
\def\[{\left[}
\def\]{\right]}

\def\<{\langle}
\def\>{\rangle}

\def\nref#1{(\ref{#1})}

\begin{document}

\thispagestyle{empty}

\renewcommand{\thefootnote}{\fnsymbol{footnote}}
\setcounter{footnote}{0}
\setcounter{figure}{0}
\begin{center}
$$$$
{\Large\textbf{\mathversion{bold} Strings in $AdS_4\times\mathbb{ CP}^3$, Wilson loops in ${\cal N}=6$ super Chern-Simons-matter and
Bremsstrahlung functions}\par}

\vspace{1.0cm}

\textrm{Jerem\'\i as Aguilera-Damia, Diego H. Correa and  Guillermo
A. Silva}
\\ \vspace{1.2cm}
\footnotesize{Instituto de F\'{\i}sica La Plata, CONICET
\\ Departamento de F\'{\i}sica, Universidad Nacional de La Plata
\\ C.C. 67, 1900 La Plata, Argentina}

\par\vspace{1.5cm}

\textbf{Abstract}\vspace{2mm}
\end{center}

\noindent

We find 1/6 BPS string configurations in $AdS_4\times\mathbb{CP}^3$,
which we identify as the duals of certain 1/6 BPS circular Wilson loops
in ${\cal N}=6$ super Chern-Simons-matter gauge theory. We use our results
to verify -in the strong coupling limit- a proposal made in
arXiv:1402.4128 for a relation  between the expectation value of these  Wilson loops
and the Bremsstrahlung function from deforming 1/2 BPS Wilson lines with a cusp.
We  also derive an analogous relation  between the expectation value
of some particular 1/12 BPS  Wilson loops and the Bremsstrahlung function from deforming
1/6 BPS Wilson lines with an internal space cusp.

\vspace*{\fill}

\setcounter{page}{1}
\renewcommand{\thefootnote}{\arabic{footnote}}
\setcounter{footnote}{0}

\newpage
\tableofcontents

\section{Introduction}

Supersymmetric Wilson loops in ${\cal N}=6$ super Chern-Simons-matter theory with gauge group $U(N)\times
U(M)$\cite{abjm,abj}, also known as ABJ theory (or ABJM when $M=N$), are constructed in terms of a generalized
$U(N|M)$ connection which includes a coupling  to the scalar and fermionic fields of the theory \cite{Drukker:2009hy}.
Such coupling is given in terms of matrices  $M^I_J$ and $\hat M^I_J$ and spinors $\eta^\alpha_I$ and $\bar\eta_\alpha^I$,
which in general depend on the parameter of the curve. Straight and circular Wilson loops, whose $M^I_J$ and $\hat M^I_J$
are constant, are among the simplest supersymmetric Wilson loops. Typical examples are the 1/6 BPS Wilson loops with $M^I_J = \hat
M^I_J ={\rm diag}(-1,1,-1,1)$ \cite{DPY,Chen:2008bp,Rey:2008bh} and 1/2 BPS Wilson loops with $M^I_J = \hat M^I_J ={\rm diag}(-1,1,1,1)$
(and certain non-vanishing $\eta$ and $\bar\eta$ in the latter) \cite{Drukker:2009hy}. Their expectation values $\langle W \rangle$
are exactly known. For the straight Wilson loops, both the 1/6 BPS and the 1/2 BPS have $\langle W \rangle = 1$, while for the circular
ones  $\langle W \rangle$ is given in terms of a matrix model \cite{Drukker:2009hy,Kapustin:2009kz,Drukker:2010nc}.

An interesting problem is to study the expectation value of some
deformations of these highly symmetric objects. Concerning the
straight Wilson loops, a natural possibility is to distort them by
adding a cusp in their trajectories. Their expectation values define
the cusp anomalous dimension, a quantity with valuable physical
interpretations \cite{PolyakovCusp,Korchemsky:1991zp}. No exact results are known for
this cusp anomalous dimension in generic situations, a notable exception is the small
angle limit for a geometrical cusp placed in the locally 1/6 BPS
\cite{Lewkowycz:2013laa}.

With respect to the circular Wilson loops, a possible generalization
is to allow $M^I_J$, $\hat M^I_J$, $\eta^\alpha_I$ and
$\bar\eta_\alpha^I$ to be specific functions of the parameter of the
curve. In particular, one can consider Wilson loops which
simultaneously move around a space-time circle and an internal space
circle. In ${\cal N}=6$ super Chern-Simons-matter theory they would
be the analogue of the ${\cal N}=4$ super Yang-Mills {\it latitude}
Wilson loops considered in \cite{Drukker:2006ga}, for which the
internal space circle is a latitude circle within a $S^2\subset S^5$
and whose radius is parametrized by an azimuthal angle $\theta_0$. In
${\cal N}=4$ super Yang-Mills theory, these 1/4 BPS latitude Wilson
loops are a particular class of loops within the larger family of
DGRT Wilson loops
\cite{Drukker:2007dw,Drukker:2007yx,Drukker:2007qr}. Latitude Wilson
loops in ${\cal N}=6$ super Chern-Simons-matter theory can be
defined as a generalization either of the 1/2 BPS
\cite{Cardinali:2012ru} or the 1/6 BPS circular Wilson loops
\cite{Marmiroli:2013nza,Bianchi:2014laa} and their vacuum
expectation values were studied perturbatively at weak coupling in
\cite{Bianchi:2014laa}.

In the case of ${\cal N}=4$ super Yang-Mills theory,  a relation
between the cusped Wilson loops vevs in the small angle limit and
the latitude Wilson loops vevs was found, which allowed the exact
computation of the Bremsstrahlung function \cite{Correa:2012at}.
With this in mind,  a similar relation was proposed for small distortions of 1/2 BPS Wilson loops in ${\cal N}=6$ super
Chern-Simons-matter theory   and tested at the first two weak coupling perturbative orders \cite{Bianchi:2014laa}.
In this regard,  one of our motivations is to further test this proposal.

In this article we study  string
configurations in $AdS_4\times\mathbb{CP}^3$, dual to latitude
Wilson loops in ${\cal N}=6$ super Chern-Simons-matter theory. We
present them and analyze their supersymmetries. We also use our
results and other considerations to verify the relation that exists
between latitude Wilson loops vevs and Bremsstrahlung functions.

\section{BPS string solutions dual to latitude Wilson loops}
\label{stringside}

In this section we study classical string configurations in $AdS_4\times\mathbb{CP}^3$ that could be
interpreted as the duals of latitude Wilson loops, {\it i.e.} circular Wilson loops whose coupling with
the scalar and fermion fields is not constant but changes along the loop. Therefore we
will focus in string configurations whose endpoints describe a circle inside $\mathbb{CP}^3$.

\subsection{Classical string configuration and supersymmetry analysis} \label{classical}

Let us begin with a presentation of the geometrical background. The
dual geometry to the ABJM theory is \cite{abjm}
\be
ds^2 = L^2\left(ds^{2}_{AdS_4} + 4ds^{2}_{\mathbb{CP}^3}\right)\,.
\label{metric}
\ee
We write the AdS  metric in global coordinates
\be
ds^{2}_{AdS_4}= -\cosh^2\!\rho\,
dt^2+d\rho^2+\sinh^2\!\rho\,\left(d\vartheta^2+\sin^2\!\vartheta\,d\psi^2\right)\,,
\label{ads}
\ee
whereas for the complex projective space one has the canonical Fubini-Study metric. An explicit expression for it can be obtained from the homogeneous coordinates ${\bf Z}=(z_1,z_2,z_3,z_4)$ parametrizing $\mathbb C^4$ \be
\begin{aligned}
z_1& =r\cos{\frac{\alpha}{2}}\cos{\frac{\theta_1}{2}}e^{\frac{i}{2}
\varphi_1}e^{\frac i4(\chi+\xi)}\,, &
z_3& =r\sin{\frac{\alpha}{2}}\cos{\frac{\theta_2}{2}}e^{\frac{i}{2} \varphi_2}e^{\frac i4(\xi-\chi)}\,,\\
z_2& =r\cos{\frac{\alpha}{2}}\sin{\frac{\theta_1}{2}}e^{-\frac{i}{2}
\varphi_1}e^{\frac i4(\chi+\xi)}\,,&
z_4& =r\sin{\frac{\alpha}{2}}\sin{\frac{\theta_2}{2}}e^{-\frac{i}{2}
\varphi_2}e^{\frac i4(\xi-\chi)}\,.
\end{aligned}
\label{comcoo}
\ee
Explicitly one has
\be
ds^2_{\mathbb C^4}=dz_I d\bar z_I=dr^2+r^2d\Omega_{7}^2\,,
\ee
with
\be
d\Omega^2_{7} = ds^{2}_{\mathbb{CP}^3} +\frac 1{16}
\left(d\xi+A\right)^2 \,,
\label{s7}
\ee
\be
A=\cos{\alpha}d\chi +
2\cos^2{\frac{\alpha}{2}}\cos{\theta_1}d\varphi_1 +
2\sin^2{\frac{\alpha}{2}}\cos{\theta_2}d\varphi_2\,.
\label{A}
\ee
The definition of $\mathbb{CP}^3$ as the equivalence relation ${\bf
Z}\sim c\, {\bf Z}$ with $ c\in {\mathbb C}^*$, amounts to `forget'
the $c=r e^{i\xi}$ coordinates in the standard $\mathbb C^4$ metric.
The result is
\begin{align}
ds^{2}_{\mathbb{CP}^3} &=
\frac{1}{4}\Bigl[d\alpha^2+\cos^2{\frac{\alpha}{2}}
\left(d\theta^2_1+\sin^2{\theta_1} d\varphi^2_1\right)+\sin^2{\frac{\alpha}{2}}\left(d\theta^2_2+\sin^2{\theta_2}d\varphi^2_2\right)\nn\\
&~~~~+\sin^2{\frac{\alpha}{2}}\cos^2{\frac{\alpha}{2}}\left(d\chi+\cos{\theta_1}d\varphi_1-\cos{\theta_2}d\varphi_2\right)^2\Bigr]\,,
\label{cp3}
\end{align}
with coordinate ranges $0\leq\alpha,\theta_1,\theta_2\leq\pi$,
$0\leq\varphi_1,\varphi_2\leq2\pi$ and $0\leq\chi\leq4\pi$.
Concomitantly \eqref{s7} manifests the statement of odd dimensional
spheres as circle bundles over projective spaces.

The geometry \eqref{metric} is supported by the following IIA fields
\be
e^{2\phi}=4\frac{L^2 }{k^2 },\qquad  F^{(4)}=\frac{3}{2}kL^2\,\mathrm{vol}({AdS_4}),\qquad  F^{(2)}=\frac{k}{4}dA\,,\\
\ee
where $\mathrm{vol}({AdS_4})=\cosh\!\rho\,\sinh^2\!\rho\,\sin\vartheta\, dt\wedge d\rho\wedge d\vartheta\wedge d\psi$.
The  curvature radius of the geometry relates to the ${\cal N}=6$ t'Hooft coupling constant $\lambda=N/k$ in the usual
way $L^2= \pi\sqrt{2\lambda}$, therefore the supergravity approximation is valid in the small curvature regime $L^4\sim\lambda\gg1$
and weak string coupling $\lambda^{5/2}/N^2\ll1$ (we have set $\alpha'=1$).

The  ABJ theory corresponds to deforming the background by turning on a $B^{(2)}$ flux over
the $\mathbb {CP}^1\subset \mathbb {CP}^3$ \cite{abj}
\be
B^{(2)}=\frac{M-N}{2k}\,  dA\,,
\label{BNS}
\ee
In \cite{abj} it was argued that unitarity is preserved if $|N-M|\le k$.

~

\begin{figure}[h]
\centering
\def\svgwidth{10cm}
 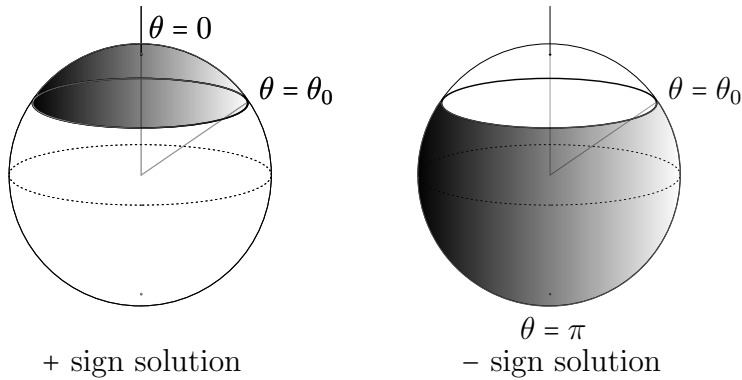
\caption{The two classical solutions}
\label{twosolutions}
\end{figure}

Our aim now is to find a string worldsheet reaching the boundary
along   a spacelike circle, while the string endpoints describe a
circle inside $\mathbb {CP}^3$. To look for the solution we start with
the Polyakov action
\be
S=\frac{1}{4\pi}\int d\tau
d\sigma\sqrt{h}h^{\alpha\beta}G_{mn}(X)\partial_{\alpha}X^{m}\partial_{\beta}X^{n}\,,
\label{act}
\ee
here $X^{m}$ represent the string coordinates, the target space metric $G_{mn}$ can be read in \eqref{metric}
and $h_{\alpha\beta}$ is an auxiliary field which implies classical equivalence between Polyakov and Nambu-Goto formulations. The appropriate ansatz
is
\be
t=0,\quad \rho=\rho(\sigma),\quad
\vartheta=\pi/2,\quad\psi=\tau,\quad \theta_1=\theta(\sigma),\quad\varphi_1=\tau,\quad\alpha=0\,.
\label{ans}
\ee
with $\tau\in(0,2\pi)$. This ansatz implies that the AdS circle will be along the equator of the $S^2$. The boundary conditions to be imposed
at infinity are
\be
\theta_1\xrightarrow[\rho\to\infty]{} \theta_0.
\ee
Plugging the ansatz into the action and fixing the conformal gauge one finds
\be
S=\frac{L^{2}}{4\pi}\int d\tau d\sigma\left[\rho'^{\,2}+\sinh^2\!\rho+\theta'^{\,2}+\sin^2\!\theta\right]\,.
\ee
The equations of motion result
\begin{align}
\rho{''}&=\sinh \rho\,\cosh\rho\,,\label{ro}\\
\theta{''}&=\sin \theta  \cos \theta  \,.
\label{eom}
\end{align}
These equations must be supplemented with the Virasoro constraints,
\be
T_{\alpha\beta}=G_{mn}(X)\partial_{\alpha}X^m\partial_{\beta}X^n-\frac{1}{2}h_{\alpha\beta}\mathcal{L} =0\,,
\ee
which in the present result in one non-trivial equation
\be
\rho'^{\,2}+\theta'^{\,2}=\sinh^2\!\rho +\sin^2\!\theta\,.
\label{vinc}
\ee
Eq. \eqref{ro} has a first integral
\be
\rho'^{\,2}=\sinh^2\!\rho +A\,.
\label{eomr}
\ee
The integration constant $A$ must be set to zero in order for the worldsheet to close smoothly in the interior of AdS and correspond to
a single loop at the boundary. From \eqref{vinc} and \eqref{eomr}  we have
\be
\theta'^{\, 2}=\sin^2\!\theta\,.
\label{eomt}
\ee
The solutions to \eqref{eomr} and \eqref{eomt} are
\begin{align}
\rho(\sigma)&=\sinh^{-1}\left(\frac{1}{\sinh\sigma}\right)\,,\nn\\
\theta(\sigma)&=\arcsin\left(\frac{1}{\cosh(\sigma_0\pm\sigma)}\right)\,,
\label{sol}
\end{align}
where we have chosen the integration constant in $\rho$ so that the range for $\sigma$ results $\sigma\in[0,\infty)$, with
the AdS boundary corresponding to $\sigma=0$. The integration constant $\sigma_0\ge0$ in \eqref{sol} sets the
boundary value $\theta_0\in(0,\frac\pi2)$
to
\be
\sin\theta_0=\frac1{\cosh\sigma_0}\,.
\label{sol2}
\ee
Note that the $\theta$ profile gives a cup-like embedding of the string in $\mathbb{CP}^3$ reaching $\theta_1=0$ or $\pi$
at the center of AdS depending on the sign chosen  in \eqref{sol} (see Figure \ref{twosolutions}).

We now proceed to evaluate the on-shell action
\begin{align}
S^{on-shell}&=\pi\sqrt{2\lambda}\int_{\sigma_{min}}^\infty d\sigma\left(\frac{1}{\sinh^2\sigma}+
\frac{1}{\cosh^2(\sigma_0\pm\sigma)}\right)\nn\\
&=\pi\sqrt{2\lambda}\left(\cosh\rho_{max}\mp\cos\theta_0 \right)\,,
\label{onshell}
\end{align}
here $\lambda$ is the ABJM 't Hooft coupling constant and we have used $L^2/\pi=\sqrt{2\lambda}$ and $\rho_{max}=\rho(\sigma_{min})$.
We have introduced  $\sigma_{min}$ in \eqref{onshell} to regulate the infinite worldsheet area, the first term in \eqref{onshell} is well understood
and known to cancel  with a boundary term, usually disregarded when writing the action, which implements
the correct boundary conditions. The final result is
\be
S^{on-shell}=\mp\pi\sqrt{2\lambda}\cos \theta_0 \,.
\label{onshell2}
\ee

~

Let us now analyze the supersymmetry of the string configuration \eqref{ans},\eqref{sol}.
We work in the Green-Schwarz formulation where the target space supersymmetries are manifest.
The fermionic partners $\Theta$ ($d=10$ Majorana spinor) of the embedding coordinates $X^m$ transform as
\be
\delta\Theta=(1+\Gamma)\kappa+\epsilon\,,
\label{fvar}
\ee
under kappa and target space supersymmetries where\footnote{$g=\dot X^2\acute X^2-(\dot X\cdot\acute X)^2$ is the
determinant of the pullback of the target space metric to the string worldvolume.}
\be
\Gamma=i\frac{\partial_\tau X^m\partial_\sigma
X^n}{\sqrt{g}}\Gamma_{mn}  \gamma_{11}\,.
\label{gamma}
\ee
The $\Gamma$ projection matrix satisfies ${\rm tr}(\Gamma)=0$ and $\Gamma^2=1$. In \eqref{fvar}, $\kappa$ is an
arbitrary local Majorana parameter and $\epsilon$ are the target space killing spinors,
which for $AdS_4\times{\mathbb{CP}}^3$ are given in the appendix \ref{killing}.

The amount of supersymmetries preserved by a given string embedding in a
particular background are the $\epsilon$ transformations which
cannot be undone by a $\kappa$ transformation and that leave the
string solution invariant. This translates into looking for solutions to
\be
(1-\Gamma)\epsilon=0\,.
\label{inv}
\ee

In what follows, we study the projection (\ref{inv}) for our string configuration.
Inserting the solution \eqref{sol} into \eqref{gamma} we obtain
\be
\Gamma = \frac{i\gamma_{11}}{\sinh^2\rho+\sin^2\theta}\left(-\sinh\rho\,\rho'\gamma_{13}+
\sinh\rho\,\theta'\gamma_{35}-\sin\theta \rho'\gamma_{17}
-\sin\theta\, \theta '\gamma_{57}\right)\,,
\label{gammacero}
\ee
while for the target space killing spinors \eqref{killesp} we get
\be
\epsilon=  \mathcal{M} \epsilon_0 = e^{\frac{\theta }{4}(\hat{\gamma}\gamma_5-\gamma_7\gamma_{11})}
e^{\frac{\rho}{2}\hat{\gamma}\gamma_1}e^{\frac{\pi}{4}\gamma_{12}}e^{\frac{\tau}{4}(-\hat{\gamma}\gamma_{11}
+\gamma_{57}+2\gamma_{23})}\epsilon_0\,.
\label{Mcero}
\ee
For analyzing \eqref{inv} and for subsequent comparison with the dual Wilson loop operator, it is useful to expand the Killing spinor $\epsilon_0$ in terms of eigenvectors of the set of matrices $\{\gamma_{01},i\hat{\gamma}\gamma_{11},i\gamma_{57},i\gamma_{49},i\gamma_{68},i\gamma_{23}\}$
with eigenvalues $\{s_0,s_1,s_2,s_3,s_4,s_5\}$ ($s_i=\pm1$) (see Appendix \ref{killing}). We write $\epsilon_0$ as
\be
\epsilon_0=\sum_{{s_i}}\theta^{(s_1,s_2,s_3,s_4)}_{(s_0,s_5)}\epsilon^{(s_1,s_2,s_3,s_4)}_{(s_0,s_5)}\,,
\label{expansion}
\ee
where $\epsilon^{(s_1,s_2,s_3,s_4)}_{(s_0,s_5)}$ and $\theta^{(s_1,s_2,s_3,s_4)}_{(s_0,s_5)}$ denotes the basis element and
the expansion coefficient respectively. Note that the included $s_1$ is redundant since $s_1=s_2s_3s_4$
\footnote{The choice of basis  $\{\hat{\gamma}\gamma_{11},\gamma_{57},\gamma_{49},\gamma_{68}\}$
is motivated by its appearance in the Killing spinor \eqref{M} associated with the phases $\xi_i$ of the embedding coordinates $z_I$.}.

Since (\ref{gammacero}) does not depend on $\tau$, the killing spinor $\tau$-dependence must be projected out.
It turns out that the appropriate projection conditions are
\begin{align}
(1+\gamma_{23}\hat{\gamma}\gamma_{11})\epsilon_0&= 0\,,\nn\\
(1-\gamma_{23}\gamma_{57})\epsilon_0&=0\,.
\label{projcond}
\end{align}
In terms of the eigenvalues $(s_1,s_2,s_3,s_4)$,  these
projections  imply $s_1=-s_2$ and by virtue of \eqref{cond} one has $s_3=-s_4$. Therefore,
the only possibilities allowed in \eqref{expansion} are: $(+,-,+,-)$, $(+,-,-,+)$, $(-,+,+,-)$ and
$(-,+,-,+)$. Notice that  \eqref{projcond} relate the $\gamma_{23}$ and $\hat{\gamma}\gamma_{11}$ eigenvalues,
 $s_5=s_1$. Therefore, the projection conditions leave
$4\times2=8$ out of the original 24 supersymmetries.

Having  imposed   (\ref{projcond}), equation (\ref{inv}) can be re-written as a condition on the constant spinor $\epsilon_0$ as
\be
(1-\mathcal{M}_{P}^{-1}\Gamma\mathcal{M}_{P})\epsilon_0=0\, ,
\label{inv2}
\ee
where $\mathcal{M}_{P}$ is $\mathcal{M}$ acting on the projected subspace.
This means that the $\tau$-dependent exponential in \eqref{Mcero} is set to one. Explicitly one obtains
\begin{align}
\mathcal{M}_{P}^{-1}\Gamma\mathcal{M}_{P} = &
\frac{i\gamma_{11}}{\sinh^2\rho+\sin^2\theta_1}\Bigl(-\sinh\rho\rho'e^{\theta_1\hat{\gamma}\gamma_5}
e^{-\frac{\pi}{2}\gamma_{12}}\gamma_{13}+\sinh\rho\theta_1'e^{\theta_1\hat{\gamma}\gamma_5}
\gamma_{35}\nn\\
&-\sin\theta_1\rho'e^{-\frac{\pi}{4}\gamma{12}}e^{\rho\hat{\gamma}\gamma_1}e^{-\frac{\pi}{4}\gamma_{12}}
\gamma_{17}-\sin\theta_1\theta_1'e^{-\frac{\pi}{4}\gamma_{12}}e^{\rho\hat{\gamma}\gamma_1}
e^{\frac{\pi}{4}\gamma_{12}}{  \gamma_{57}}\Bigr)
\nn\\
= & i\gamma_{11}(\sin\theta_0\gamma_{27} - \cos\theta_0\gamma_{57})\,.
\label{tilde}
\end{align}
where, in the final line, explicit solution (\ref{sol}) has been used.
Note that this operator is coordinate independent and unaffected by the sign choice in \eqref{sol}, which means
that both classical configurations preserve the same supersymmetries.
Since the operator in \eqref{tilde} commutes with the projection conditions
\eqref{projcond} they can be simultaneously diagonalized.  The outcome is that only
half of the eigenvectors of $\mathcal{M}_{P}^{-1}\Gamma\mathcal{M}_{P}$ satisfy
(\ref{inv2}), leaving 4 conserved supercharges. Thus we conclude that configuration \eqref{sol} is 1/6 BPS.

Projections  \eqref{projcond} imply that $s_1=s_5=-s_2$, which results in 8 independent components.
If we further impose \eqref{inv2} we get the conditions
\be
\sin\theta_0\theta^{(s_2,-s_2,s_3,-s_3)}_{(s_0,s_2)} -s_0(1-s_0s_2\cos\theta_0)\theta^{(-s_2,s_2,s_3,-s_3)}_{(s_0,-s_2)}
= 0\,, \label{explicit}\ee
which can be solved as follows, using $\nu=\cos\theta_0$,  in terms of four independent coefficients
\begin{alignat}{4}
\theta^{(+-+-)}_{(++)} & =  \sqrt{1-\nu} \, \omega_1\,, & \qquad
\theta^{(-++-)}_{(+-)}& = \sqrt{1+\nu} \,\omega_1\,,
\nn\\
\theta^{(+-+-)}_{(-+)} & =\sqrt{1+\nu} \,\omega_2\,,  & \qquad
\theta^{(-++-)}_{(--)}& = -\sqrt{1-\nu} \,\omega_2\,,
\nn\\
\theta^{(+--+)}_{(++)}& =\sqrt{1-\nu} \,\omega_3\,,  & \qquad
\theta^{(-+-+)}_{(+-)} & = \sqrt{1+\nu} \, \omega_3\,,
\nn\\
\theta^{(+--+)}_{(-+)} & =\sqrt{1+\nu} \,\omega_4\,,  & \qquad
\theta^{(-+-+)}_{(--)}& = -\sqrt{1-\nu} \, \omega_4\,.
\label{susycuerda}
\end{alignat}

As an aside, note that in the Killing spinors (\ref{Mcero}) we have set $\theta_2=0$ and $\varphi_2=0$.
However, since the sphere spanned by $\theta_2$ and  $\varphi_2$ is shrunk to zero
size, se should be able to keep them arbitrary and preserve the same supersymmetries.
Consider for example taking $\theta_2=\theta_1$ and $\varphi_2=\varphi_1$ and still having
$\alpha =0$.  While the $\Gamma$ projector remains as (\ref{gammacero}), the corresponding Killing spinor
would be defined by
\be
\mathcal{M} = e^{\frac{\theta_1}{4}(\hat{\gamma}\gamma_5-\gamma_7\gamma_{11} + \gamma_{98}+\gamma_{46})}
e^{\frac{\rho}{2}\hat{\gamma}\gamma_1}
e^{\frac{\pi}{4}\gamma_{12}}e^{\frac{\tau}{2}\gamma_{23}}
e^{-\frac{\tau}{4}(\hat{\gamma}\gamma_{11}-\gamma_{57}+\gamma_{49}-\gamma_{68})}\,.
\label{Mcero2}
\ee
At first sight, this may appear problematic since the $\tau$-dependence
cannot be projected out from $\mathcal{M}$. Nevertheless this is not a problem,
since what it matters is to project out the $\tau$-dependence from
$\mathcal{M}^{-1}\Gamma\mathcal{M}$. Since $\gamma_{98}$, $\gamma_{46}$,
$\gamma_{49}$ and $\gamma_{68}$ commutes with $\Gamma$, upon
imposing (\ref{projcond}) one finds that
$\mathcal{M}_{P}^{-1}\Gamma\mathcal{M}_{P}$ is given by \eqref{tilde}, either
for $\mathcal{M}$ defined in \nref{Mcero} or for $\mathcal{M}$
defined in \nref{Mcero2}. Therefore, the kappa symmetry equation is
not modified leading to the preservation of the same
supersymmetries.

\subsection{Dual Wilson loop operators}
As we will see in this section, the previous semiclassical string
configuration is dual to a kind of BPS latitude Wilson loop.
The term {\it latitude} was used in \cite{Bianchi:2014laa}
to refer to a deformation of circular Wilson loops that involves both, a
geometrical azimuth on the $ S^2\subset AdS_4$ and an internal
space azimuth on some $S^2 \subset \mathbb{CP}^3$. It was observed
nevertheless that their expectation values depend on a single combination
of the two azimuths: $\nu = \sin\theta_{geo}\cos\theta_{int}$.  For the sake
of simplicity, we set the geometrical circle at the equator, {\it i.e.}
$\theta_{geo} =\frac{\pi}{2}$, and call $\theta_0$ the internal space azimuth
 $\theta_{int}$.

BPS Wilson loops in ${\cal N}=6$ super Chern-Simons-matter theory have been constructed in terms of a
$U(N|M)$ connection \cite{Drukker:2009hy,Cardinali:2012ru}
\be
L =\left(\begin{array}{cc}
        A_\mu\dot{x}^\mu -\frac{2\pi i}{k}|\dot x|M^I_{J}C_I \bar C^J  & -i\sqrt{\frac{2\pi}{k}}|\dot x|\eta_I^\alpha \bar\psi^I_\alpha \\
        -i\sqrt{\frac{2\pi}{k}}|\dot x|\psi_I^\alpha \bar\eta^I_\alpha & \hat A_\mu\dot{x}^\mu +\frac{2\pi i}{k}|\dot x|\hat M^I_{J}
        \bar C^J C_I
      \end{array}
\right)\,, \label{connection} \ee
as
\be
W_F = \frac{1}{{\cal N}_{\cal T}}{\rm STr}\left[P e^{i\oint_{\cal C} L d\tau}{\cal T}\right]\,,
\label{halfBPS}
\ee
where  ${\cal N}_{\cal T} = {\rm STr({\cal T})}$ is a normalization
factor and ${\cal T}$ is a twisting matrix which depends on the
particular choice of $M^I_J, \hat M^I_J, \eta_I^\alpha$ and
$\bar\eta^I_\alpha$, which is necessary for the Wilson loop to be
gauge invariant.

We are interested in identifying the Wilson loop operator dual to the string configuration of section \ref{stringside}.
Therefore, we will consider the contour ${\cal C}$ in \eqref{halfBPS} to be the unit circle $\vec x(\tau) =  (0,\cos\tau,\sin\tau)$. In this section we will
identify the specific choice of $M^I_J, \hat M^I_J, \eta_I^\alpha$ and $\bar\eta^I_\alpha$. We shall start with $M^I_J = \hat M^I_J$
in the ABJM case, {\it i.e.} gauge group ranks $M=N$,  which has a neater geometrical interpretation \cite{Rey:2008bh}.
For the kind of BPS Wilson loop we are interested we take
\be
M^I_J = \hat M^I_J = \delta^I_J -\frac{2
\dot{z}_J\dot{\bar{z}}^I}{|\dot{z}|^2}\,,
\label{Mfromz}
\ee
where $z_I(\tau)$ is the trajectory of the endpoints of the string
configuration inside ${\mathbb{CP}^3}$, expressed in terms of the
complex coordinates given in \eqref{comcoo}. For the classical
string solution \eqref{ans},\eqref{sol} we have
\be z_1 = \cos\tfrac{\theta_0}{2} e^{i\frac{\tau}{2}}\,,\quad z_2 =
\sin\tfrac{\theta_0}{2} e^{-i\frac{\tau}{2}}\,,\quad z_3 = 0\,,\quad
z_4 = 0\,, \ee
which leads to
\be
M^I_J = \hat M^I_J = \left(
  \begin{array}{cccc}
    -\nu & e^{-i\tau}\sqrt{1-\nu^2} & 0 & 0 \\
    e^{i\tau}\sqrt{1-\nu^2} & \nu & 0 & 0 \\
    0 & 0 & 1 & 0 \\
    0 & 0 & 0 & 1 \\
  \end{array}
\right)\,,
\label{Mmatrix}
\ee
This matrices, altogether with  spinor couplings given by
\be
\eta^{\alpha}_I=\frac{e^{\frac{i\nu\tau}{2}}}{\sqrt{2}}\left(
  \begin{array}{c}
    \sqrt{1+\nu} \\
    -\sqrt{1-\nu}e^{i\tau} \\
    0 \\
    0 \\
  \end{array}\right)_I
  \,\left(
      1,\ -ie^{-i\tau}
  \right)^{\alpha}\,,\qquad \bar{\eta}^{\alpha}_I=i(\eta^{\alpha}_I)^{\dag}\,,
\label{eta}
\ee
give rise to a family of 1/6 BPS Wilson loops. Their supersymmetry parameters
$\bar\Theta^{IJ} = \bar\theta^{IJ} - (x\cdot\gamma)\bar\epsilon^{IJ}$, which has been
explicitly spelled out in \cite{Bianchi:2014laa}, are such that\footnote{$\bar{\theta}^{IJ}_{\alpha}$ and $\bar{\epsilon}^{IJ}_{\alpha}$
generates super Poincar\'e and super conformal transformations respectively, where  $\alpha$ is a spinor index and $IJ$ are antisymmetrized $SU(4)$ indices in the fundamental
representation.}
\begin{alignat}{4}
   \zeta^{13}_{-1} &=\tfrac{1}{2}(\bar{\theta}^{13}_1 - i\bar{\epsilon}^{13}_1) = \sqrt{1-\nu}\, \omega_1\,, & \qquad
   \zeta^{23}_{-2} &=\tfrac{1}{2}(i\bar{\theta}^{23}_2 - \bar{\epsilon}^{23}_2) = \sqrt{1+\nu}\, \omega_1\,,
   \nn\\
    \zeta^{13}_{+1}&=\tfrac{1}{2}(\bar{\theta}^{13}_1 + i\bar{\epsilon}^{13}_1) = \sqrt{1+\nu}\, \omega_2\,, & \qquad
    \zeta^{23}_{+2}&=\tfrac{1}{2}(i\bar{\theta}^{23}_2 + \bar{\epsilon}^{23}_2) = \sqrt{1-\nu}\, \omega_2\,,
   \nn\\
   \zeta^{14}_{-1}&=\tfrac{1}{2}(\bar{\theta}^{14}_1 - i\bar{\epsilon}^{14}_1) = \sqrt{1-\nu}\, \omega_3\,, & \qquad
    \zeta^{24}_{-2}&=\tfrac{1}{2}(i\bar{\theta}^{24}_2 - \bar{\epsilon}^{24}_2) = \sqrt{1+\nu}\, \omega_3\,,
   \nn\\
    \zeta^{14}_{+1}&=\tfrac{1}{2}(\bar{\theta}^{14}_1 + i\bar{\epsilon}^{14}_1) = \sqrt{1+\nu}\, \omega_4\,, & \qquad
   \zeta^{24}_{+2}&=\tfrac{1}{2}(i\bar{\theta}^{24}_2 + \bar{\epsilon}^{24}_2) = \sqrt{1-\nu}\, \omega_4\,,
\label{wlsusy}
\end{alignat}
Note that these supercharges  coincide exactly with \eqref{susycuerda}, provided
the identification between $\zeta^{IJ}_{\pm\alpha}$ and $\theta^{(s_1,s_2,s-3,s_4)}_{(s_0,s_5)}$
given in the Appendix \eqref{argument} is used.

~

Let us conclude this section studying a family of bosonic Wilson loops, also considered in
\cite{Bianchi:2014laa}, that correspond to a latitude deformation of the well known bosonic
1/6 BPS circular Wilson loop\footnote{Analogously one can define
a $U(M)$ Wilson loop as
\be
\hat W_B = \frac{1}{M}{\rm Tr}\left[P e^{i\oint
\left(\hat A_{\mu}\dot{x}^{\mu}-\frac{2\pi i}{k}|\dot{x}|\hat M^{I}_J
\bar{C}^{J}C_I\right)d\tau }\right]\,,\nn
\ee
where $\hat M^{I}_J = M^{I}_J$. }
\be
W_B = \frac{1}{N}{\rm Tr}\left[P e^{i\oint
\left(A_{\mu}\dot{x}^{\mu}-\frac{2\pi i}{k}|\dot{x}|M^{I}_J
C_I\bar{C}^{J}\right)d\tau }\right]\,,
\label{doceBPS}
\ee
where
\be
M^I_J =
\left(
  \begin{array}{cccc}
    -\nu & e^{-i\tau}\sqrt{1-\nu^2} & 0 & 0 \\
    e^{i\tau}\sqrt{1-\nu^2} & \nu & 0 & 0 \\
    0 & 0 & -1 & 0 \\
    0 & 0 & 0 & 1 \\
  \end{array}
\right)\,.
\label{Mdoce}
\ee
The supercharges preserved by this bosonic Wilson loop happen to be
a subset of the supercharges given by \eqref{wlsusy}. More specifically they are
obtained by setting $\omega_1=\omega_4=0$ in \eqref{wlsusy}, leaving 2 free
parameters and then concluding that this bosonic Wilson loops are 1/12 BPS.

We would like to analyze whether there is a relation between the dual of the 1/6 BPS
latitude Wilson loop and the dual of the bosonic 1/12 BPS latitude Wilson loop in terms
of a geometrical smearing, as it is the case for the 1/2 BPS Wilson loop
and the bosonic 1/6 BPS Wilson loop (see \cite{DPY}). Recall that the scalar coupling of the latter,
$M^I_J = \hat M^I_J = {\rm diag}(-1,1,-1,1)$, cannot be realized
as \eqref{Mfromz} for any $z_I(\tau)$\footnote{A matrix given by $\delta^I_J -\frac{2\dot{z}_J\dot{\bar{z}}^I}{|\dot{z}|^2}$ will always
have eigenvalues $\{-1,1,1,1\}$.},
the suggestion in \cite{DPY} was to interpret the bosonic Wilson loop \eqref{doceBPS} with
$M^I_J = \hat M^I_J = {\rm diag}(-1,1,-1,1)$ not as dual to a single string configuration
but as a the dual to a distribution of strings smeared over a $\mathbb{CP}^1\subset\mathbb{CP}^3$.
The amount  supersymmetry  preserved by the smearing is understood as follows:
if one considers rotations of  string configurations dual to the 1/2 BPS Wilson loop along
the aforementioned $\mathbb{CP}^1$ some of the supercharges will depend
on the angles of the rotations. The supersymmetries of the smeared
distribution are only those supercharges independent of the rotation angles,
which are precisely the supercharges of the bosonic 1/6 BPS Wilson loop \cite{DPY}.

The bosonic Wilson loop defined with $M^I_J$ given by \eqref{Mdoce},
which has eigenvalues  $\{-1,1,-1,1\}$,  cannot correspond to a single
string either. Since by turning off the latitude deformations setting $\nu =1$ we have the relation
described in the previous paragraph, we would like to analyze what happen if one smears over a
$\mathbb{CP}^1$  strings dual to 1/6 BPS latitude Wilson loops. More specifically we would like to ask
whether there are common supercharges among the rotated configurations. For this purpose, we construct a 2-parameter family of string configurations
related to the one of section \ref{classical} via a $SU(4)$ rotation on the
$\mathbb{CP}^3$ embedding coordinates.

Writing
$${\bf Z}=(z_1,z_2,z_3,z_4)=(\vec z,\vec w)$$
with $\vec z=(z_1,z_2)$ and $\vec w=(z_3,z_4)$, the solution found
on the previous section having $\alpha=0$ corresponds to
$$\vec z_0=(\cos\frac {\theta(\sigma)}2e^{i\frac\tau2},\sin\frac{\theta(\sigma)}2e^{-i\frac\tau2}),~~\vec w_0=0$$
Acting on it with the following $SU(2)$ element
$$g(\alpha_0,\phi_0)=\left(
\begin{array}{ll}
~~~\cos\frac{\alpha_0}2 & -\sin\frac{\alpha_0}2e^{i\frac{\phi_0}2}\\
\sin\frac{\alpha_0}2e^{-i\frac{\phi_0}2} & ~~~\cos\frac{\alpha_0}2
                           \end{array}
                           \right)$$
one finds
\be
{\bf Z}=(\vec z_0,0)\to{\bf Z}'=(\cos\frac{\alpha_0}2 \vec z_0,\sin\frac{\alpha_0}2e^{-i\frac{\phi_0}2}\vec z_0)
\label{rota}
\ee
It is straight forward to see that this rotated configuration
satisfies the classical equations of motion. The new solution reads
\be t=0,~
\rho=\rho(\sigma),~\vartheta=\pi/2,~\psi=\tau,~\alpha=\alpha_0,~\theta_1=\theta_2=\theta(\sigma),~\varphi_1=\varphi_2=\tau,~\chi=\phi_0
\label{ans2}
\ee
Since we have obtained the solution acting with a symmetry of the
action, the value of the on-shell action does not change.

The supersymmetry analysis for these configurations is made in
Appendix \ref{kappa2}, where we find that the killing equation has
the same form as \eqref{inv2} but in a rotated base of spinors.
Therefore, they preserve the same amount of supersymmetry, {\it
i.e.} they are all 1/6 BPS. However there is no common subspace of
solutions for the kappa symmetry equation between the different
configurations parametrized by $(\alpha_0,\phi_0)$. Therefore,
a smeared configuration obtained from the rotations defined in \eqref{rota},
cannot be regarded as the dual of any BPS Wilson loop. In particular it would not
correspond to the dual of the 1/12 BPS bosonic Wilson loop defined by
\eqref{doceBPS}-\eqref{Mdoce}.

Given the fact that the preserved supersymmetries of the 1/12 BPS bosonic Wilson loop
\eqref{doceBPS}-\eqref{Mdoce} are a subset of the preserved supersymmetries of the 1/6
BPS latitude Wilson loop, it can still be possible that the dual of the former
is interpreted as some more general smearing of the dual of the latter. To further
speculate about this possibility let us note that a projection that would enforce
$\omega_1=\omega_4=0$ would require to set $s_0-s_3=0$ in \eqref{susycuerda}.  This
condition is clearly equivalent to imposing the projection
\be
(1-i\gamma_{01}\gamma_{49})\epsilon_0=0\,.
\label{smearing}
\ee
However, at the moment we do not have an interpretation of \eqref{smearing} in terms of a geometrical smearing.
Note that such a projection that relates $s_0$ and $s_3$ cannot be obtained as a consequence
of smearing with rotations acting on $\mathbb{CP}^3$ only.

\section{Bremsstrahlung functions and latitude Wilson loops}
\label{relationwithB}
One of our motivations to study latitude Wilson loops is the possibility of relating their expectation
values with Bremsstrahlung functions, as it is the case in ${\cal N}=4$ super Yang-Mills theory \cite{Correa:2012at}. The prospect
of such a relation in ${\cal N}=6$ super Chern-Simons-matter theory has also been considered in \cite{Bianchi:2014laa}. We will
now further elaborate on this possibility.

The Bremsstrahlung functions are related to the expectation values of straight Wilson loops with small cusps.
If one considers a line in $\mathbb{R}^3$ with a cusp at some point
\be
\langle W_{\rm cusp} \rangle = e^{-\Gamma_{\rm cusp}\log\frac{\Lambda_{IR}}{\Lambda_{UV}}}\,,
\ee
where $\Lambda_{IR}$ and $\Lambda_{UV}$ are infrared and ultraviolet cutoffs respectively \cite{PolyakovCusp,Korchemsky:1991zp}. Given that we could distort
either a 1/2 BPS straight Wilson loop or 1/6 BPS straight Wilson loop with cusps, we shall distinguish between $B_{1/2}$ and $B_{1/6}$ Bremsstrahlung functions accordingly.

Moreover, in each of the cases it is possible to distort the
straight Wilson loop with either a geometrical cusp angle $\phi$ or
an internal cusp angle $\theta$. Since a 1/2 BPS straight Wilson
loop distorted with two cusp angles such that $\theta = \pm \phi$
remains BPS, one has a unique Bremsstrahlung function $B_{1/2}$.
Therefore, when $\theta,\phi\ll 1$,
\be
\Gamma_{\rm cusp} = (\theta^2-\phi^2) B_{1/2}(\lambda)\,.
\ee

However, a 1/6 BPS straight Wilson loop distorted with two cusp angles is not BPS,  not even for $\theta = \pm\phi$.
Therefore, we have to distinguish between internal and geometrical cusp angles Bremsstrahlung functions. For
$\theta,\phi\ll 1$,
\be
\Gamma_{\rm cusp} = \theta^2 B^{\theta}_{1/6}(\lambda)-\phi^2  B^{\phi}_{1/6}(\lambda)\,.
\ee

We now analyze the relation between these Bremsstrahlung functions and the latitude Wilson loop we have been studying. In \cite{Bianchi:2014laa}, the proposal
\be B_{1/2}(\lambda) = \frac{1}{4\pi^2}\left.\frac{\partial}{\partial\nu}\log \langle W_F \rangle \right|_{\nu=1}\,,
\label{rela12}
\ee
was check up to two-loop in the weak coupling expansion, with $\langle W_F \rangle$  computed at framing 0. Since the relation \eqref{rela12}
has not been derived or proven, verifying that it is also satisfied
in the strong coupling limit can be  seen as compelling evidence
that it may be valid to all-loop order.  This Bremsstrahlung function
has been computed in the strong coupling limit from a classical string ending in a cusped line
in \cite{forini}, obtaining to leading order the result
\be
B_{1/2}=\frac{\sqrt{2\lambda}}{4\pi} + {\cal O}(1)\,.
\label{lhs}
\ee
To test \eqref{rela12} in this limit we need $\langle W_F \rangle$, which
at leading order is
\be
\langle W_F \rangle = e^{-S^{on-shell}} + {\cal O}(1) = e^{\pi\sqrt{2\lambda}\nu} + {\cal O}(1)
\label{lead}
\ee
where the on-shell action has been evaluated in \eqref{onshell2}. We have
chosen the sign that minimizes the action and dominates the saddle
point approximation. Upon using \eqref{lead} to compute r.h.s. of \eqref{rela12}
we observe the agreement with \eqref{lhs}.

Let us now turn to the other Bremsstrahlung functions. Concerning $B^\phi_{1/6}(\lambda)$, it has
been noted in \cite{Bianchi:2014laa} that a relation analogous to \eqref{rela12} would fail already
at leading order in the weak coupling expansion. In passing, we would like to mention that there
exists nevertheless a proposed exact expression
for $B^\phi_{1/6}(\lambda)$ in terms of the derivatives of a multiply wound Wilson loop \cite{Lewkowycz:2013laa},
but we will not discuss here.

On the other hand, the analogous relation for $B^\theta_{1/6}(\lambda)$,
\be
B^\theta_{1/6}(\lambda) = \frac{1}{4\pi^2}\left.\frac{\partial}{\partial\nu}\log \langle W_B \rangle \right|_{\nu=1}\,,
\label{rela16}
\ee
can be checked to leading weak coupling order with the two-loop results of \cite{Griguolo:2012iq} and \cite{Bianchi:2014laa}.
By means of an analysis similar to the one in \cite{Correa:2012at}, we will now argue that \eqref{rela16} is valid to all-loop order.

We will consider a bosonic Wilson loop with internal cusp angle $\theta$ which is of the form
\be
W_c = \frac{1}{N}{\rm Tr}\left[P e^{i\oint_{{\cal C}_1+{\cal C}_2}
\left(A_{\mu}\dot{x}^{\mu}-\frac{2\pi i}{k}|\dot{x}|{M_c}^{I}_J
C_I\bar{C}^{J}\right)d\tau }\right]\,, \label{cusp}
\ee
where ${\cal C}_1$ and ${\cal C}_2$ are two radial lines in $\mathbb{R}^3$. There is no geometrical cusp between
the lines but the coupling with scalar fields changes from ${\cal C}_1$ to ${\cal C}_2$
\be
M_c=\left\{\begin{array}{cc}
                   M(0) & {\rm if\ } x(\tau)\in {\cal C}_1\,, \\
                   M(\theta)
                    & {\rm if\ } x(\tau)\in {\cal C}_2\,, \\
                 \end{array}\right.
\ee
with
\be
M(\theta) = \left(
  \begin{array}{cccc}
    -\cos\theta & -\sin\theta & 0 & 0 \\
    -\sin\theta & \cos\theta & 0 & 0 \\
    0 & 0 & -1 & 0 \\
    0 & 0 & 0 & 1 \\
  \end{array}
\right)\,.
\ee
We will parametrize the half-lines with the logarithm of the radial distance, which is related to the global time when mapping   $\mathbb{R}^3$ to $\mathbb{R}\times S^2$.
For instance, for the half-line ${\cal C}_2$ we use
$x^\mu = (e^\tau,0,0)$ for which $\int_{-\infty}^{\infty} d\tau \sim \Delta T = \log\frac{\Lambda_{IR}}{\Lambda_{UV}}$.
Expanding for small values of the internal cusp angle $\theta$ we obtain to leading order
\begin{align}
\langle\delta W_c\rangle &= \langle W_c\rangle-\langle W_c\rangle_{\theta=0} = - \theta^2 B^\theta_{1/6} \log\frac{\Lambda_{IR}}{\Lambda_{UV}}
\nn\\
&= \frac{{\theta}^2}{2} \left(\frac{2\pi}{k}\right)^2 \int_{{\cal
C}_2}d\tau_1\int_{{\cal C}_2}d\tau_2 ({m_c})^I_J({m_c})^K_L
e^{\tau_2} e^{\tau_1} \langle\!\langle
{\Phi(\tau_1)}^J_I{\Phi(\tau_2)}^L_K\rangle\!\rangle_{\rm
straight}\,,
\label{expcusp}
\end{align}
where ${\phi(\tau)}^J_I=C(x(\tau))_I{\bar{C}(x(\tau))}^J$ is an operator in the adjoint of $U(N)$ and ${m_c}$ comes from the
first order expansion of the matrix $M_c$,
\be {m_c}=\left(
  \begin{array}{cccc}
    0 & -1 & 0 & 0 \\
    -1 & 0 & 0 & 0 \\
    0 & 0 & 0 & 0 \\
    0 & 0 & 0 & 0 \\
  \end{array}
\right)\,.
\ee
The double brackets denote correlation functions along the Wilson loop (with no cusp).
In general we can define them for any Wilson loop as
\be
\langle\!\langle \mathcal{O}(\tau_1)\mathcal{O}(\tau_2)\rangle\!\rangle_{\cal C}
=\frac{
\langle{\rm Tr}
\left[P \mathcal{O}(\tau_1) \mathcal{O}(\tau_2)
e^{i\oint_{\cal C} \left(A_{\mu}\dot{x}^{\mu}-\frac{2\pi i}{k}|\dot{x}|M^{I}_J
C_I\bar{C}^{J}\right)d\tau }\right]\rangle}
{
\langle{\rm Tr}
\left[P e^{i\oint_{\cal C} \left(A_{\mu}\dot{x}^{\mu}-\frac{2\pi i}{k}|\dot{x}|M^{I}_J
C_I\bar{C}^{J}\right)d\tau }\right]\rangle}
\label{doble}
\ee

The structure of the double brackets, as correlation functions of in a 1-dimensional
theory, are constrained by conformal symmetry. When writing \eqref{expcusp} we
have already used that one-point double brackets are vanishing. In the present case, 2-point double brackets
are determined up to an overall constant $\gamma$ (see Appendix \ref{prospa})
\be
\langle\!\langle {\phi(\tau_1)}^J_I{\phi(\tau_2)}^L_K\rangle\!\rangle_{\rm straight} =
\frac{\gamma e^{-\tau_1} e^{-\tau_2} \delta^{J}_K\delta^L_I }{2(\cosh (\tau_1-\tau_2)-1)}
\label{cuspint}\,,\qquad\left(I,J,K,L=1,2\right)\,.
\ee
Inserting \eqref{cuspint} in \eqref{expcusp} and eliminating one of the integrals
as $\Delta T = \log\frac{\Lambda_{IR}}{\Lambda_{UV}}$, we obtain
\be B_{1/6}^\theta = -\frac{2\pi^2\gamma}{k^2}
\int_{-\infty}^{+\infty}\frac{d\tau}{\cosh\tau-1}= \frac{4\pi^2
\gamma}{k^2} \label{B16theta} \ee
where the integral was regularized and a UV divergence was discarded.

We have related directly the Bremsstrahlung function $B_{1/6}^\theta$ with
the coefficient $\gamma$ in the double bracket two-point correlator, defined
with the straight 1/6 BPS Wilson loop. Now, by a similar argument
we relate the derivative of the latitude Wilson loop vev with
the coefficient $\gamma$ in the double bracket two-point correlator
for the circular 1/6 BPS Wilson loop.

We start by considering a latitude Wilson loop with a very small azimuth $\theta_0$ and compute
\begin{align}
{\theta_0}^2\frac{\partial \log\langle
W_B\rangle\mid_{\theta_0=0}}{\partial{\theta_0}^2}
&\simeq \frac{\langle W_B\rangle -  \left.\langle W_B\rangle\right|_{\theta_0=0}}
{\left.\langle W_B\rangle\right|_{\theta_0=0}}\nn\\
&\simeq
\frac{{\theta_0}^2}{2}\left(\frac{2\pi}{k}\right)^2\int_{\cal C}
d\tau_1\int_{\cal C} d\tau_2 {m(\tau_1)}^I_J{m(\tau_2)}^K_L
\langle\!\langle {\phi(\tau_1)}^J_I{\phi(\tau_2)}^L_K\rangle\!\rangle_{\rm circle}\,,
\label{expcirc}
\end{align}
where ${\cal C}$ is a unit circle and the matrix $m(\tau)$ is given by
\be m(\tau)=\left(
  \begin{array}{cccc}
    0 & e^{-\tau} & 0 & 0 \\
    e^{\tau} & 0 & 0 & 0 \\
    0 & 0 & 0 & 0 \\
    0 & 0 & 0 & 0 \\
  \end{array}
\right)\,.
\ee
Now the integration in the double bracket is over the circular
contour ${\cal C}$, for which (see Appendix \ref{prospa})
\be
\langle\!\langle {\Phi(\tau_1)}^J_I{\Phi(\tau_2)}^L_K\rangle\!\rangle_{\rm circle} =
\frac{\gamma \delta^J_K\delta^L_I}{2(1-\cos (\tau_1-\tau_2))}\,,\qquad\left(I,J,K,L=1,2\right)\,.
\label{circint}
\ee
Note that since the straight and the circular Wilson loops are
related by a conformal transformation and \eqref{doble} is conformal
invariant, the coefficient $\gamma$ appearing in \eqref{circint} is
the same as the one in \eqref{cuspint}. Inserting \eqref{circint}
into \eqref{expcirc} we obtain
\be \frac{\partial \log\langle
W_B\rangle\mid_{\theta_0=0}}{\partial{\theta_0}^2} =
\frac{1}{4}\left(\frac{2\pi}{k}\right)^2\gamma
\int_0^{2\pi}d\tau_1\int_0^{2\pi}d\tau_2
\frac{\cos(\tau_1-\tau_2)}{1-\cos(\tau_1-\tau_2)}=
- \frac{8\pi^4\gamma}{k^2}\,,
\label{circ}
\ee
where again a UV  was eliminated through regularization. If we now compare
with \eqref{B16theta} we conclude that
\be
B_{1/6}^{\theta}=-\frac{1}{2\pi^2}\frac{\partial}{\partial{\theta_0}^2}\log\langle
W_B\rangle\mid_{\theta_0=0}\,,
\label{final}
\ee
or in terms of the parameter $\nu=\cos\theta_0$
\be
B_{1/6}^{\theta}=\frac{1}{4\pi^2}\frac{\partial}{\partial\nu}\log\langle
W_B\rangle\mid_{\nu=1} \,.
\label{final2}
\ee
%

\section{Conclusions}
We have studied latitude Wilson loops in ${\cal N}=6$ super Chern-Simons-matter theory
and their relation to Bremsstrahlung functions. By latitude Wilson loops we mean certain class of
circular Wilson loops, whose coupling with the scalar and fermion fields changes along an internal
space circle as the position in the geometrical space-time circle changes. They are
generalizations of either the 1/2 BPS or the 1/6 BPS circular Wilson loops.

More specifically we have studied the description of such latitude Wilson loops in the strong coupling limit, in terms of classical
strings in the type IIA background $AdS_{4}\times\mathbb{CP}^3$. We have found a  family of 1/6 BPS classical string
solutions that we have identified  with the 1/6 BPS latitude Wilson loops discussed in \cite{Bianchi:2014laa}.
Our string solutions are the analogues of the 1/4 BPS circular ones found in $AdS_5\times S^5$ \cite{Drukker:2006ga}. As
in the ${\cal N}=4$ SYM case, the strong coupling limit for  the latitude Wilson loops vevs can be obtained from the circular Wilson loop vev by
the replacement $\lambda\to\lambda\cos^2\theta_0$. However, it is known that this relation is not valid to all orders in $\lambda$, in
particular it is violated in the weak coupling limit \cite{Bianchi:2014laa}. This prevents from finding a simple relation between the Bremsstrahlung
function and $\lambda$-derivatives of the circular Wilson loop, which vev can be computed from a matrix model \cite{Drukker:2010nc}.

Concerning the bosonic 1/12 BPS latitude Wilson loops given by \eqref{doceBPS}-\eqref{Mdoce}, they cannot be described in the strong coupling limit
by a single string because its coupling matrix $M^I_J$ cannot be represented in the form of \eqref{Mfromz}. It would be then interesting to further
investigate if they can be described in terms of a geometrical smearing of 1/6 BPS latitude strings. As we discussed in
the text, smearing only in the internal space $\mathbb{CP}^3$ does not work, in contrast to the case of bosonic 1/6 BPS \cite{DPY}.

In \cite{Bianchi:2014laa} a relation between the Bremsstrahlung function associated with the cusp deformation of 1/2 a BPS Wilson line
and derivatives of the latitude Wilson loop has been proposed. We have verified such proposal, which had been verified in the weak coupling limit in \cite{Bianchi:2014laa},
in the strong coupling regime. This is compelling evidence that the relation \eqref{rela12} should be valid to all-loop order.

Moreover, we have derived the expression  \eqref{rela16}  for the Bremsstrahlung function associated with an internal cusp
deformation of the 1/6 BPS Wilson line in terms of derivatives of the bosonic 1/12 BPS latitude Wilson loops \eqref{doceBPS}-\eqref{Mdoce}.
In this case, the derivation is similar to the one presented in \cite{Correa:2012at} for the $\mathcal{N}=4$ SYM Bremsstrahlung function,
which relies on the conformal symmetry of the problem.

Another interesting problem to consider in the future is to analyze if a similar derivation can be provided for the relation proposed in \cite{Bianchi:2014laa}.
Also, in order to make this kind of relations between the Bremsstrahlung functions and latitude Wilson loops more useful, it would be important
to investigate whether the latter can be computed exactly by some supersymmetric localization argument.

~

{\bf Acknowledgements }

We would like to thank G.Giribet,  M.Leoni and J.Maldacena for discussions.
This work was supported by CONICET and grants PICT 2010-0724, PICT 2012-0417,
PIP 0681 and PIP 0396.

\appendix

\section{$AdS_{4}\times\mathbb{CP}^3$ Killing spinors} \label{killing}

Target space $AdS_4\times \mathbb{CP}^3$ can be found from the (maximally supersymmetric) 11-dimensional supergravity solution
$AdS_4\times S^7$ via a Kaluza-Klein reduction. Thus, $AdS_{4}\times\mathbb{CP}^3$ Killing spinors are a subset
of those of $AdS_{4}\times S^7$. Killing spinors in $d=11$ are given by the solutions to
\be
\nabla_\mu\epsilon+\frac{1}{288}(\Gamma^{\nu\rho\sigma\tau}_{\mu}-8\delta^{\nu}_{\mu}\Gamma^{\rho\sigma\tau})F_{\nu\rho\sigma\tau}=0\,,
\label{gravitino}
\ee
where $\nabla_\mu$ is the standard covariant derivative containing the spin connection and
$\mu$ runs over all the 11 coordinates. We denote
tangent space gamma matrices as $\gamma^a = e^a_\mu \Gamma^\mu$, with the
following elfbeine basis
\begin{align}
e^0 &=L\cosh\rho\,dt\,, & e^1 &= L\,d\rho\,, & e^2 &= L\sinh\rho\,d\vartheta\,,\nn\\
e^3 &=L\sinh\rho\sin\vartheta\,d\psi\,, & e^4 &= L  d\a, & e^5 &= L  \cos\frac{\a}{2}\,d\theta_1,\nn\\
e^6 &= L  \sin\frac{\a}{2} \,d\theta_2, & e^7 &= L  \cos\frac{\a}{2} \sin\theta_1\,d\varphi_1, & e^8 &= L \sin\frac{\a}{2}\sin\theta_2 \, d\varphi_2\,, \nn\\
e^{11} &=-\frac L2\left(d\zeta+A\right) &
e^9 &= L  \cos\frac{\a}{2}\sin\frac{\a}{2} \Bigl(d\chi+ \cos \theta_1 \, d\varphi_1 - \cos\theta_2\,d\varphi_2  \Bigr),\hspace{-4cm}
\end{align}
$A$ was defined in \eqref{A}.

The 4-form in the $d=11$ solution is simply proportional to the $AdS$ volume form,
$F_{\mu\nu\rho\sigma}=6\,\varepsilon_{\mu\nu\rho\sigma}$, reducing \eqref{gravitino}
to the Killing spinor equation
\be
\nabla_{\mu}\epsilon=\frac{1}{2}\hat{\gamma}\Gamma_\mu\epsilon\,,
\label{killeq}
\ee
here $\hat{\gamma}=\gamma^0\gamma^1\gamma^2\gamma^3$. The solution to
\eqref{killesp} can be written as \cite{DPY}
\be
\epsilon(x)=\mathcal{M}(x)\,\epsilon_0\,,
\label{killesp}
\ee
where
\begin{align}
\mathcal{M}(x)&=
e^{\frac{\alpha}{4}(\hat{\gamma}\gamma_4-\gamma_9\gamma_{11})}e^{\frac{\theta_1}{4}(\hat{\gamma}\gamma_5-\gamma_7\gamma_{11})}
e^{\frac{\theta_2}{4}(\gamma_{98}+\gamma_{46})}e^{-\frac{\xi_1}{2}\hat{\gamma}\gamma_{11}}e^{-\frac{\xi_2}{2}\gamma_{57}}\nn\\
&\,\cdot\; e^{-\frac{\xi_3}{2}\gamma_{49}}e^{-\frac{\xi_4}{2}\gamma_{68}}
e^{\frac{\rho}{2}\hat{\gamma}\gamma_1}e^{\frac{t}{2}\hat{\gamma}\gamma_0}e^{\frac{\vartheta}{2}\gamma_{12}}e^{\frac{\psi}{2}\gamma_{23}}\,,
\label{M}
\end{align}
with
\be
\xi_1=\frac{2\varphi_1+\chi+\xi}{4}, \quad\,
\xi_2=\frac{-2\varphi_1+\chi+\xi}{4},\quad\,
\xi_3=\frac{2\varphi_2-\chi+\xi}{4}, \quad\,
\xi_4=\frac{-2\varphi_2-\chi+\xi}{4}\,.
\ee
In \eqref{killesp} the constant spinor $\epsilon_0$ has 32 real components and all $\gamma$'s in \eqref{M}
are flat. Since all the matrices multiplying the phases $\xi_i$ in \eqref{M}: $i\hat{\gamma}\gamma_{11}$, $i\gamma_{57}$,
$i\gamma_{49}$ and $i\gamma_{68}$ are traceless, square to the identity and commute among themselves, we choose
$\epsilon_0$ to be an eigenvector of the set
\be
i\hat{\gamma}\gamma_{11}\epsilon_0=s_1\epsilon_0,~~i\gamma_{57}\epsilon_0=s_2\epsilon_0,
~~i\gamma_{49}\epsilon_0=s_3\epsilon_0,~~i\gamma_{68}\epsilon_0=s_4\epsilon_0\,,
\label{ss}
\ee
where all $s_i$ are $\pm 1$. Note that these matrices are not all independent because in odd dimensions the
product of all gamma matrices gives the identity matrix
\be
\hat{\gamma}\gamma_{11}\gamma_{57}\gamma_{49}\gamma_{68}=\gamma_{0123456789}\gamma_{11}=\pm1\,.
\label{chiral}
\ee
Choosing our set of gamma matrices to satisfy $\gamma_{0123456789}\gamma_{11}=+1$, we see that
there are only three independent eigenvalues among \eqref{ss}: the eigenvalues must satisfy
$s_1s_2s_3s_4=1$. This leaves us with the following possibilities for the $\epsilon_0$ eigenvalues
\begin{gather}
(+,+,+,+),(+,+,-,-),(+,-,-,+),(+,-,+,-),\nn\\
(-,+,-,+),(-,+,+,-),(-,-,+,+),(-,-,-,-),\nn
\end{gather}
Each of these choices corresponds to four independent spinors which could be further classified
in terms of the eigenvalues of $\gamma_{01}$ and $i\gamma_{23}$.  Generically we will write the spinor $\epsilon_0$
as in \eqref{expansion}.

The reduction to ten dimensions is accomplished  along the $\xi$
direction. Therefore, to find the IIA Killing spinors we demand
invariance under $\xi\rightarrow\xi+\delta\xi$. This shift results
in
\be \epsilon(x)\to
\epsilon^{\prime}(x)=\mathcal{M}(x)\,e^{{\frac{i\delta\xi}{8}}(i\hat{\gamma}\gamma_{11}+i\gamma_{57}+i\gamma_{49}+i\gamma_{68})}\epsilon_0\,.
\label{xitrasl} \ee
Thus, invariance under $\delta\xi$ in \eqref{xitrasl} translates into
\be
s_1+s_2+s_3+s_4=0\,.
\label{cond}
\ee
This condition eliminates the $(+,+,+,+)$ and $(-,-,-,-)$ cases and implies that $AdS_4\times\mathbb{CP}^3$ preserves 3/4 of the original
32 supersymmetries, this means 24 supercharges\footnote{This same analysis shows that the $AdS_4\times S^7/{\mathbb Z_k}$ solution preserves
also 24 supersymmetries except for the $k=1,2$ cases.}.

\section{Supersymmetry correspondence}\label{argument}

In our supersymmetry analysis  we use the following  representation for the  $\gamma$ matrices
\be
\begin{array}{lll}
    \gamma_{0}=i\sigma_2\otimes \mathbb{I} \otimes \mathbb{I} \otimes \mathbb{I} \otimes \mathbb{I}\,, & &
    \gamma_{1}=\sigma_1\otimes \mathbb{I} \otimes \mathbb{I} \otimes \mathbb{I} \otimes \mathbb{I}\,, \\
    \gamma_{2}=\sigma_3\otimes \sigma_2 \otimes \mathbb{I} \otimes \mathbb{I} \otimes \mathbb{I}\,, & &
    \gamma_{3}=\sigma_3\otimes \sigma_1 \otimes \mathbb{I} \otimes \mathbb{I} \otimes \mathbb{I}\,, \\
    \gamma_{4}=\sigma_3\otimes \sigma_3 \otimes \sigma_3 \otimes \sigma_2 \otimes \mathbb{I}\,, & &
    \gamma_{5}=\sigma_3\otimes \sigma_3 \otimes \sigma_2 \otimes \mathbb{I} \otimes \mathbb{I}\,, \\
    \gamma_{6}=\sigma_3\otimes \sigma_3 \otimes \sigma_3 \otimes \sigma_3 \otimes \sigma_2\,, & &
    \gamma_{7}=\sigma_3\otimes \sigma_3 \otimes \sigma_1 \otimes \mathbb{I} \otimes \mathbb{I}\,, \\
    \gamma_{8}=\sigma_3\otimes \sigma_3 \otimes \sigma_3 \otimes \sigma_3 \otimes \sigma_1\,, & &
    \gamma_{9}=\sigma_3\otimes \sigma_3 \otimes \sigma_3 \otimes \sigma_1 \otimes \mathbb{I}\,,
  \end{array}
\ee
for which ${\gamma_{01}}$, $\gamma_{23}$, $\gamma_{57}$, $\gamma_{49}$ and $\gamma_{68}$ are diagonal.

We want to identify the preserved supercharges of the latitude Wilson loops \eqref{wlsusy}
with the preserved supercharges of the string configuration \eqref{susycuerda}. To begin with,
we should understand how the bulk space quantum numbers $s_i$ are related to antisymmetric pairs of $SU(4)$ indices $I,J$.

Recall that the $su(4)$ Lie algebra generators  $R^{I}_{J}$, in the fundamental representation, act as follows
\be
R^I_J\,|z^K\rangle = \delta_{J}^{K}\,|z^{I}\rangle-\frac{1}{4}\delta^{I}_{J}\,|z^{K}\rangle\,.
\ee
The operators $R^1_1$, $R^2_2$ and $R^3_3$ commute among themselves and can be
identified with the 3-dimensional Cartan subalgebra of $su(4)$\footnote{Note that the $R^4_4$ operator is not independent
since $R^1_1+R^2_2+R^3_3+R^4_4=0$.}. The $R^I_I$ operators have a diagonal form
\be
R^1_1 =\left(
        \begin{array}{cccc}
          {\frac{3}{4}}  \\
           & {-\frac{1}{4}}  \\
          & & {-\frac{1}{4}}  \\
          && & {-\frac{1}{4}}  \\
        \end{array}
      \right) , \quad
      R^2_2=\left(
        \begin{array}{cccc}
          {-\frac{1}{4}}   \\
           & {\frac{3}{4}}  \\
          & & {-\frac{1}{4}}  \\
          && & {-\frac{1}{4}}  \\
        \end{array}
      \right) , \quad
      R^3_3=\left(
        \begin{array}{cccc}
          {-\frac{1}{4}}   \\
           & {-\frac{1}{4}}   \\
          & & {\frac{3}{4}} \\
          && & {-\frac{1}{4}} \\
        \end{array}
      \right)\,.
\ee
By inspecting the action of the generators $R^1_1$, $R^2_2$, $R^3_3$ and $R^4_4$ on the projective space coordinates $z_I$, one
realizes that they induce shifts in the phases  $\xi_1$,  $\xi_2$,  $\xi_3$ and  $\xi_4$ respectively, which motivates the following
identification
\be
\{R^1_1,R^2_2,R^3_3,R^4_4\}\longleftrightarrow
\{i\hat{\gamma}\gamma_{11},i\gamma_{57},i\gamma_{49},i\gamma_{68}\}\,.
\label{ident2}
\ee
Therefore every $\zeta^{IJ}$ can be identified with a specific
$\zeta^{(s_1,s_2,s_3,s_4)}$. For instance $\zeta^{12}\leftrightarrow\zeta^{(+,+,-,-)}$,
$\zeta^{13}\leftrightarrow\zeta^{(+,-,+,-)}$, etc. Essentially, $s_I$ and $s_J$ are taken positive, while
the other two are taken negative.

The bulk quantum numbers $(s_0,s_5)$ can also be identified 3-dimensional spinorial indices. For
the conventions used in \cite{Bianchi:2014laa} these identifications are as follows.
\be
\begin{array}{lcrclcr}
\zeta^{13}_{+1}&\longleftrightarrow& \theta^{(+-+-)}_{(-+)}\,,&\qquad  &\zeta^{13}_{-1}&\longleftrightarrow& \theta^{(+-+-)}_{(++)}\,,\nn\\
\zeta^{23}_{+2}&\longleftrightarrow& -\theta^{(-++-)}_{(--)}\,,&\qquad &\zeta^{23}_{-2}&\longleftrightarrow& \theta^{(-++-)}_{(+-)}\,,\nn\\
\zeta^{14}_{+1}&\longleftrightarrow&  \theta^{(+--+)}_{(-+)}\,,&\qquad &\zeta^{14}_{-1}&\longleftrightarrow& \theta^{(+--+)}_{(++)}\,,\nn\\
\zeta^{24}_{+2}&\longleftrightarrow& -\theta^{(-+-+)}_{(--)}\,,&\qquad
&\zeta^{24}_{-2}&\longleftrightarrow& \theta^{(-+-+)}_{(+-)}\,.
\label{ident}
\end{array}
\ee

\section{Supersymmetry of the rotated solutions}\label{kappa2}

We would like to analyze the condition (\ref{inv2}) again, this time for the solution with the extra two parameters $(\alpha_0,\phi_0)$  given in \eqref{ans2}.
We need to write $\Gamma$ and the matrix $\mathcal{M}$ defining the Killing spinors. Inserting the solution \eqref{ans2} into
\eqref{gamma} we obtain
\begin{align}
\Gamma'&=\frac{i}{\sinh^2\rho+\sin^2\theta}\nn\\
&\left[\rho'\,\sinh\rho\,\gamma_{31}+
\theta'\sinh\rho\,\gamma_{3}\left
(\cos\frac{\alpha_0}2\gamma_5+\sin\frac{\alpha_0}2\gamma_6\right)
+\rho'\sin\theta\left (\cos\frac{\alpha_0}2\gamma_7+\sin\frac{\alpha_0}2\gamma_8\right)\,\gamma_{1}\right.\nn\\
&~~\left. +\theta'\sin\theta\left
(\cos\frac{\alpha_0}2\gamma_7+\sin\frac{\alpha_0}2\gamma_8\right)
\left
(\cos\frac{\alpha_0}2\gamma_5+\sin\frac{\alpha_0}2\gamma_6\right)
\right]\gamma_{11}\,.
\label{gammaev}
\end{align}
In comparison with the $\Gamma$ given in \nref{gammacero}, there is an extra $\alpha_0$ dependence. The $\alpha_0$ dependence can be
factorized in terms of a rotation in the planes 56 and 78 of the tangent space.
\be
\Gamma'=e^{-a/2}\,\Gamma\,e^{a/2}\,,\qquad a=\frac{\alpha_0}2\left(\gamma_{56}+\gamma_{78}\right)\,.
\ee
The matrix $\mathcal{M}'$  after the rotation takes the form
\begin{align}
\mathcal{M}'&=
e^{\frac{\alpha_0}{4}(\hat{\gamma}\gamma_4-\gamma_9\gamma_{11})}
e^{\frac{\theta}{4}(\hat{\gamma}\gamma_5-\gamma_7\gamma_{11}+\gamma_{98}+\gamma_{46})}
e^{-\frac{\phi_0}{8}(\hat{\gamma}\gamma_{11}+\gamma_{57}-\gamma_{49}-\gamma_{68})}\nn\\
&\cdot~
e^{-\frac{\tau}{4}(\hat{\gamma}\gamma_{11}-\gamma_{57}+\gamma_{49}-\gamma_{68})}
e^{\frac{\rho}{2}\hat{\gamma}\gamma_1}
e^{\frac{\pi}{4}\gamma_{12}}e^{\frac{\tau}{2}\gamma_{23}}
\nn\\
&= e^{\frac{\alpha_0}{4}(\hat{\gamma}\gamma_4-\gamma_9\gamma_{11})}
\mathcal{M}
e^{-\frac{\phi_0}{8}(\hat{\gamma}\gamma_{11}+\gamma_{57}-\gamma_{49}-\gamma_{68})}\,,
\end{align}
where $\mathcal{M}$ is the one defined in \nref{Mcero2}.

In order to consider $\mathcal{M'}^{-1}\Gamma'\mathcal{M'}$, it is convenient to collect the two exponentials depending on $\alpha_0$ in a single rotation $R$,
\be
R:=e^{\frac{\alpha_0}{4}(\hat{\gamma}\gamma_4-\gamma_9\gamma_{11}+\gamma_{56}+\gamma_{78})}\,,
\label{rot}
\ee
and define the rotated matrices as
\be
\tilde{A} = R A R^{-1}\,.
\ee
For example, for the rotated gamma matrices we obtain,
\begin{alignat}{4}
\tilde\gamma_{4} &= \cos\frac{\alpha}{2}\gamma_4 +
\sin\frac{\alpha}{2} \hat{\gamma}\,, \quad & \tilde{\hat{\gamma}} &=
\cos\frac{\alpha}{2}\hat{\gamma} - \sin\frac{\alpha}{2}\gamma_4\,,
\nn\\
\tilde\gamma_{9} &= \cos\frac{\alpha}{2}\gamma_9 +
\sin\frac{\alpha}{2}\gamma_{11}\,, \quad & \tilde\gamma_{11} &=
\cos\frac{\alpha}{2}\gamma_{11} - \sin\frac{\alpha}{2}\gamma_{9}\,,
\nn\\
\tilde\gamma_{5} &= \cos\frac{\alpha}{2}\gamma_5 -
\sin\frac{\alpha}{2}\gamma_6 \quad & \tilde\gamma_{6}\,,
&=\cos\frac{\alpha}{2}\gamma_6 + \sin\frac{\alpha}{2}\gamma_5\,,
\nn\\
\tilde\gamma_{7} &= \cos\frac{\alpha}{2}\gamma_7 -
\sin\frac{\alpha}{2}\gamma_8\,, \quad & \tilde\gamma_{8}
&=\cos\frac{\alpha}{2}\gamma_8 + \sin\frac{\alpha}{2}\gamma_7\,,
\end{alignat}
In what follows, it will be important that the following
combinations of gamma matrices remain invariant under the rotation
\begin{alignat}{2}
\tilde{\hat{\gamma}}\tilde \gamma_{{11}} +\gamma_{\tilde{49}} &=
\hat{\gamma}\gamma_{11}+\gamma_{49}\,,
\nn\\
\tilde\gamma_{{57}}+\tilde\gamma_{68} &= \gamma_{57}+\gamma_{68}\,,
 \nn\\
\tilde\hat{\gamma}\tilde\gamma_{5}+\tilde\gamma_{46}&=\hat{\gamma}\gamma_{5}+\gamma_{46}\,,
\nn\\
-\tilde\gamma_{7}\tilde\gamma_{11}+\tilde\gamma_{98}&=-\gamma_{7}\gamma_{11}+\gamma_{98}\,,
\end{alignat}
which imply that
\be
\tilde{\cal M} = R {\cal M}R^{-1} = {\cal M}\,.
\ee
We can then conclude that
\be
\mathcal{M'}_P^{-1}\Gamma'\mathcal{M'}_P =
e^{\frac{\phi_0}{8}(\hat{\gamma}\gamma_{11}+\gamma_{57}-\gamma_{49}-\gamma_{68})}R^{-1}\mathcal{M}_{P}^{-1}\Gamma\mathcal{M}_{P}
R
e^{-\frac{\phi_0}{8}(\hat{\gamma}\gamma_{11}+\gamma_{57}-\gamma_{49}-\gamma_{68})}\,.
\ee
Therefore, the SUSY equation in the rotated base is the same as in
the previous case ($\alpha=0$) and we can conclude that this
configuration is 1/6 BPS too.

Note that the corresponding base of killing eigenvectors are
parametrized by the $\alpha_0$ value that defines the rotation
(\ref{rot}) in the spinor space. So, even though the amount of
preserved supersymmetries is always the same, each configuration
with different $\alpha_0$ values preserves a different set of them.
We can search for the common set of eigenvectors between all of
these different bases. This common set is the subspace that remains
invariant under the action of (\ref{rot}). In other words, we are
searching for the solutions of
\be
\left(\hat\gamma\gamma_{4}-\gamma_{9}\gamma_{11}+\gamma_{56}+\gamma_{78}\right)\epsilon_0=0\,.
\label{smear}
\ee
Making use of these conditions we rewrite (\ref{smear}) to the form
\be
(1-s_2s_4)(-\gamma_{9}\gamma_{11}+\gamma_{78})\epsilon_0=0\,.
\ee
The last equation is satisfied only by spinors that satisfy
$(1+\gamma_{57}\gamma_{68})\epsilon_0=0$. This projection does not
commute with conditions \eqref{projcond}, and from \eqref{explicit}
it is straight forward seeing that both projections do not have a
common space of solutions.

\section{CFT correlators in projective space coordinates}\label{prospa}
In this section we review how CFT correlation functions can be written in terms of coordinates
of a higher dimensional projective space \cite{Weinberg:2010fx}. The group of conformal transformations in a $d$-dimensional
space-time can be realized in terms of rotations in a $d+2$-dimensional projective space. For a $d=3$ Euclidean space the
conformal group is $SO(1,4)$, so we will work with the cone defined by
\be
X\cdot X = \eta_{AB} X^A X^B =0\,,
\ee
where $A,B=1,2,\dots 5$ and $\eta_{AB} = {\rm diag}(1,1,1,1,-1)$. Since $X^A$ are coordinates of a
projective space $c X^A$ and $X^A$ should be identified for any non-vanishing $c$. We can relate space-time coordinates
$x^\mu$ ($\mu=1,2,3$) with the projective space ones according to
\be
x^\mu = \frac{X^\mu}{X^4+X^5}\,,
\label{projective}
\ee
so that conformal transformations acting on the $x^\mu$ are simply $SO(1,4)$ rotations acting on $X^A$. With these definitions is not difficult to see that,
\be
X \cdot X' = -\frac{1}{2}(X^4+X^5)({X'}^4+{X'}^5)(x-x')^2\,.
\ee
Tensor fields in the projective space are then related to tensor fields in the 3-dimensional space. In particular, a space-time scalar
field $\phi$ of conformal dimension $\Delta$ relates to a $SO(1,4)$ scalar field $\Phi$ according to
\be
\phi(x) = (X^4+X^5)^\Delta \Phi(X)\,.
\ee
Therefore, for a pair of scalar fields of equal conformal dimension we have that
\begin{align}
\langle \phi_1(x)\phi_2(x')\rangle &= (X^4+X^5)^\Delta ({X'}^4+{X'}^5)^\Delta \langle \Phi_1(X)\Phi_2(X')\rangle =
\frac{(X^4+X^4)^\Delta ({X'}^4+{X'}^5)^\Delta}{(-2 X\cdot X')^\Delta}\nn\\
& = \frac{1}{(x-x')^{2\Delta}}\,.
\end{align}
In the last equation we have been referring to ordinary vacuum expectation values.
However, conformal symmetry of the problem also constrain the two-point double bracket correlator
defined in \eqref{doble}. Either for the straight or the circular Wilson loop we have
\be
\langle\!\langle \phi^I_J(x(\tau)) \phi^K_L(x(\tau'))
\rangle\!\rangle = \gamma(\lambda)\ \frac{(X^4(\tau)+X^5(\tau))
(X^4(\tau')+X^5(\tau'))}{-2 X(\tau)\cdot X(\tau')}
\delta^I_L\delta^K_J\,,
\label{generica}
\ee
where $\phi^I_J(x) = C_J(x)\bar C^I(x)$ and $I,J,K,L$ is understood as taking the values 1,2 hereafter.
Note that in \eqref{generica} the only $\lambda$-dependent comes through $\gamma$,
{\it i.e.} no anomalous dimension develops. A key point for this asseveration is that the insertion
should preserve some of the Wilson loop supersymmetries. This is precisely the case for the insertions
$C_1(x)\bar C^2(x)$ and $C_2(x)\bar C^1(x)$ considered in \eqref{expcusp},\eqref{expcirc}
when the Wilson loop has $M^I_J = {\rm diag}(-1,1,-1,1)$\footnote{It can be seen that
$C_3(x)\bar C^4(x)$ and $C_4(x)\bar C^3(x)$ also satisfy this condition.}.

Let us now evaluate \eqref{generica} for a half-line and a circle. We parametrize a half-line in  $\mathbb{R}^3$ as
\be
(x^1,x^2,x^3) = (e^{\tau},0,0)\,, \qquad\tau\in(-\infty,\infty)\,,
\ee
where $\tau$ is in correspondence with Euclidean time in $\mathbb{R}\times S^2$.
In terms of projective coordinates \eqref{projective} the curve reads
\be
(X^1,X^2,X^3,X^4,X^5) = (1,0,0,-\sinh\tau,\cosh\tau)\,,
\ee
and then from \eqref{generica} one gets
\be
\langle\!\langle \phi^I_J(x(\tau)) \phi^K_L(x(\tau')) \rangle\!\rangle_{\rm straight} =
 \frac{\gamma e^{-\tau} e^{-\tau'} \delta^I_L\delta^K_J}{2\cosh(\tau -\tau')-2}\,,
\ee

For the circular loop in $\mathbb{R}^3$
\be
(x^1,x^2,x^3) = (0,\cos\tau,\sin\tau)\,,
\ee
we can use
\be
(X^1,X^2,X^3,X^4,X^5) =  (0,\cos\tau,\sin\tau,0,1)\,,
\ee
and then
\be
\langle\!\langle \phi^I_J(x(\tau)) \phi^K_L(x(\tau')) \rangle\!\rangle_{\rm circle} =
 \frac{\gamma \delta^I_L\delta^K_J}{2-2\cos(\tau -\tau')}\,.
\ee


\end{document}